# The influence of neutron star mergers on the galactic chemical enrichment of r-process elements.

*E. De Donder and D. Vanbeveren*

*Astrophysical Institute, Vrije Universiteit Brussel, Pleinlaan 2, 1050 Brussels*

E-mail: ededonde@vub.ac.be
dvbevere@vub.ac.be

## Abstract

A population number synthesis code follows in detail the evolution of a population of single stars and of close binaries. We use our code to simulate the population of neutron star - neutron star and black hole - neutron star binaries. We then combine our population number synthesis code with a galactic chemical evolutionary model in order to follow the time evolution of the formation and merger rate of these double compact star binaries and the resulting chemical enrichment of r-process elements, over the whole Galactic lifetime. It can be concluded that the neutron star/black hole merger process is able to reproduce the observed r-process enrichment of the Galaxy. However, we show that the latter conclusion depends critically on the physics of case BB Roche lobe overflow in binaries with a neutron star component and a hydrogen deficient core helium/helium shell burning star with a mass between 2.6 $M_\odot$ and 6 $M_\odot$.

## 1. Introduction

The idea that the rapid neutron-capture process (r-process) is responsible for the existence of the heaviest elements in the universe was realised some time ago (Suess and Urey, 1956; Burbidge et al., 1957; Cameron, 1957). Cowan et al. (1991) reviewed the possible astrophysical sites where the r-process can happen. Two sites have favourable physical conditions in order to be a major r-process source: the supernova (SN) explosion of a massive star (SN II, SN Ib/c) (Woosley et al., 1994; Takahashi et al., 1994; Hoffman et al., 1997; Qian and Woosley, 1996; Meyer et al., 1998) and the binary neutron star merger (NSM) (Davies et al., 1994; Janka and Ruffert, 1996; Baumgarte et al., 1997; Ruffert and Janka, 1998; Rosswog et al., 1999, 2000, 2001; Freiburghaus et al., 1999a, b). A discussion of the pro's and the contra's of both sites is given in Qian (2000) and Rosswog et al. (2001). At present, neither of the two can be promoted as the main enrichment source without reasonable doubt.

Abundances of r-process elements have been observed in halo and disk stars covering a metallicity range [Fe/H] ~ -3.1 – 0.5 (e.g. Woolf et al., 1995; Shetrone, 1996; Sneden et al., 2000; Burris et al., 2000; Cayrel et al., 2001; Hill et al., 2002). Figure 8 shows the temporal evolution of the element europium of which the behaviour can be considered as typical for the r-process elements. From the observed [r/Fe][1] versus [Fe/H] relation the following main features are noticed:

- [r/Fe] is on average ≥ 0 for –2.5 ≤ [Fe/H] ≤ 0 and increases with decreasing metallicity.
- a large scatter is present in the range –3.1 ≤ [Fe/H] ≤ -1 that goes up to a factor of 1000 at very low metallicity
- a few very iron-poor ([Fe/H] ≈ -(3.1-2.9)) halo giants (CS 31082-001, CS 22892-052) are extremely rich in r-process elements (up to a factor of 40-50 times the solar value)

[1] $[r/Fe]=\log(r/Fe)_{\star} - \log(r/Fe)_{\odot}$

The presence of r-process elements in very metal-poor halo stars indicates that the enrichment in r-process elements must have started already at the beginning of the Galaxy. The large scatter is interpreted as the result of incomplete mixing of the interstellar medium with the ejecta of a very rare event (like the NSM event) much rarer than the SNII and SNIbc occurrence rate.

Theoretical predictions of the [r/Fe] versus [Fe/H] relation with a chemical evolutionary model (CEM) have been made by several groups (e.g. Andreani et al., 1988; Mathews et al., 1992; Pagel and Tautvaisiene, 1997; Travaglio et al., 1999, 2001; Ishimaru and Wanajo, 1999). They all promote SNe resulting from single stars with an initial mass in the range 8-10 $M_{\odot}$ as the major enrichers of r-process elements. Ishimaru and Wanajo (1999) added also the explosion of ≥ 30 $M_{\odot}$ stars to explain the large dispersion in [r/Fe] in halo stars. Binary neutron star mergers were only considered in the study of Mathews et al. (1992), however in a far too qualitative way and without the use of detailed binary evolutionary computations.

When a double NS binary (NS+NS) forms, its further spiral-in evolution is governed by angular momentum and orbital energy loss via gravitational wave radiation. It is well known from general relativity that the timescale on which the binary merges and thus ejects the r-elements, sensitively depends on the total mass, orbital period and eccentricity of the system. Therefore to make reliable predictions on the merger rate of double NS binaries, the physical properties of the systems at the time of their formation must be well known. It is the scope of the present paper to combine a population number synthesis (PNS) code with a CEM to compute the time evolution of the double NS binary population, their merger rate and the abundance ratio [r/Fe] resulting from the latter. We also include the population of black hole - neutron star binaries (BH+NS) which may in the same way as merging double NS binaries produce r-elements. In particular, we will investigate whether or not it is possible that NSMs only can explain the observed galactic enrichment of r-process elements.

In section 2 we summarise our PNS code with special attention for all the processes that affect the theoretical prediction of the considered binary population. Section 3 deals with the CEM and the link with PNS. The results are presented in section 4.

## 2. The PNS model

A PNS code with single stars and interacting binaries has to account for:

- the evolution of single stars
- the evolution of case A, case $B_r$, case $B_c$ and case C binaries (as defined by Kippenhahn & Weigert (1967) and Lauterborn (1969)) with a mass ratio q > 0.2 (q is the mass of the secondary divided by the mass of the primary and the primary is always defined as the originally most massive component), accounting in detail for the effects of Roche lobe overflow (RLOF), mass transfer, mass and angular momentum loss from the system and common envelope (CE) evolution
- the evolution of binaries with a small mass ratio (q ≤ 0.2) which is governed by the spiral-in process
- the evolution of mergers for which we make a distinction between the merging of two normal stars, a normal star and a compact star and of two compact stars.
- the effects of a (asymmetric) supernova (SN) explosion on the binary parameters
- a detailed treatment of the evolution of the binary orbital period, which depends critically on the physics of the previously listed processes

Our PNS code uses the following distribution functions to initialise the stellar parameters and to model the asymmetry of a SN explosion:

- the initial mass function (IMF): we use a Salpeter type distribution (Salpeter 1955) for single and binary primary masses. Notice that the conclusions of the present paper do not critically depend on the adopted IMF.

- the binary mass ratio distribution $\phi(q)$: we make our simulations for a distribution according which the number of systems increases as q decreases (Hogeveen, 1991, 1992; we will further use the term 'Hogeveen distribution'), for a flat distribution, and for a distribution that peaks at q=1 (Garmany et al., 1980)
- the initial binary period distribution is taken flat in the log(P) (Popova et al. 1982; Vereshchagin et al. 1987, 1988) with P between $P_{min}$[2] and 10 years. Most of the binaries with a period larger than 10 years will not interact and their components are treated as single stars.
- asymmetric supernova explosions induce neutron star kicks, which drastically affect the binary periods. We assume that the direction of the SN kicks is isotropic and that they have magnitudes that follow a $\chi^2$-like distribution $f(v_{kick})$, corresponding with the observed space velocity distribution of single pulsars from Lorimer et al. (1997). Accounting for possible observational uncertainties, we calculate the survival probability of the binary for two average kick velocities: $\langle v_{kick} \rangle$ = 150 km/s and 450 km/s. i.e.

$$f(v_{kick}) = 1.96 \cdot 10^{-6} v_{kick}^{3/2} e^{-v_{kick}/171}$$
$$\text{and} \qquad (1)$$
$$f(v_{kick}) = 2.7 \cdot 10^{-5} v_{kick}^{3/2} e^{-v_{kick}/60}$$

The velocity distribution of isolated radio pulsars has been re-analysed recently by Arzoumanian et al. (2002). They promote a two-component distribution with characteristic velocities of 90 km/s and 500 km/s. Test calculations illustrate that the PNS results of NS+NS and BH+NS binaries with the latter distribution are somewhere in between the results holding for the two distributions given by (1).

Our PNS code has been used earlier in order to study the supernova rates, the single and binary WR, O-type star population and the double compact star binary population in regions of continuous star formation (De Donder and Vanbeveren, 1998 hereafter DDV; Vanbeveren et al., 1998a, b, c). De Donder and Vanbeveren (2002, 2003) added the evolution of intermediate mass single stars and binaries in order to study the time evolution of the Galactic SN rates including the SN Ia's. The code relies on a large set of stellar evolutionary computations of intermediate mass and massive single stars and binary components of which the evolution is followed up to the end of their core helium burning (CHeB) phase, with a moderate amount of convective core overshooting and for the metallicity range $0.001 \le Z \le 0.02$. The evolutionary tracks of massive stars are computed with the most recent stellar wind mass-loss rate ($\dot{M}$) formalisms. It is mainly the mass-loss rate during the hydrogen deficient CHeB phase and its uncertainties that significantly affect the results of the present paper. For this reason we briefly discuss the rates in section 2.1.

[2] $P_{min}$ is the required minimum period to avoid contact of both components on the zero-age main sequence (ZAMS). For most of the binaries, $P_{min} \approx 1$ day.

One of the crucial parameters in binary evolution is the amount of matter lost by the mass loser due to RLOF and that is (or can be) accreted by the companion star (originally introduced by Vanbeveren et al. (1979) as the parameter β) and the accompanying loss of orbital angular momentum when mass leaves the binary. We discuss both shortly in subsection 2.3.

### 2.1. The stellar wind mass-loss rate formalism in massive stars during the hydrogen deficient CHeB phase.

A massive star becomes a hydrogen deficient CHeB star after extensive mass loss during the Luminous Blue Variable (LBV) phase, the Red Supergiant (RSG) phase or the RLOF phase. Observable counterparts are the Wolf-Rayet (WR) stars and the O-type subdwarfs.

Our preferred WR $\dot{M}$–formalism that is used in our stellar evolutionary code, has been discussed in detail in Vanbeveren et al. (1998a, b, c) and in Van Bever and Vanbeveren (2003), i.e.

$$\log(\dot{M}) = \log L - 10 + 0.5\log(X_{Fe}/X_{Fe,solar}). \qquad (2)$$

with $X_{Fe}$ the iron abundance. Assuming that the WR SW is radiation driven, we expect that the heavy elements (primarily iron) are the main wind drivers and thus that $\dot{M}$ mainly depends on the iron abundance.

The population of double compact star binaries obviously depend on the pre core collapse (≈ end CHeB) and post core collapse stellar masses. They are given in figure 1 and 2. We remark that our masses are much larger than those published before 1997 by different research groups. With our preferred WR mass-loss rate formalism, Galactic stars with initial mass between 40 $M_{\odot}$ and 100 $M_{\odot}$ end their life with a mass between 10 $M_{\odot}$ and 20 $M_{\odot}$ corresponding to carbon-oxygen (CO) cores masses between 5 $M_{\odot}$ and 15 $M_{\odot}$.

Figure 1 and 2 also show the pre- and post core collapse masses of massive stars in the Small Magellanic Cloud (SMC) ($Z = 0.1Z_{\odot}$), assuming that the WR mass-loss rate scales proportional to $\sqrt{X_{Fe}}$.

The WR stars which are observed in binaries have a mass > 5 $M_{\odot}$. However, as will be demonstrated here, the NS merger results depend critically on the evolution of the hydrogen deficient CHeB stars with a mass ≤ 5 $M_{\odot}$. Question: does equation (1) also applies in the lower mass range? In section 2.5 we will discuss the possible evolutionary effect of stellar wind mass loss during the CHeB phase of hydrogen deficient stars with a mass ≤ 5 $M_{\odot}$ and the consequences for PNS of NSMs.

### 2.2. Black hole formation

As outlined in more detail in De Donder and Vanbeveren (2002, 2003), when we link our CHeB evolutionary calculations (previous subsection) with the post-CHeB tracks of Woosley and Weaver (1995) we conclude that massive BHs (mass larger than 4-5 $M_{\odot}$, up to 10 $M_{\odot}$ and even larger) are formed from progenitors with an initial mass ≥ 40 $M_{\odot}$. Lower mass BHs (mass between 2 $M_{\odot}$ and 4-5 $M_{\odot}$) are descendants from massive single stars with an initial mass between ~25 $M_{\odot}$ and ~ 40 $M_{\odot}$.

Some of the LMXB-BH candidates and Cyg X-1 have large space velocities which may be an indication that SN-like mass ejection occurred prior to BH formation (Nelemans et al. 1999). The α-elements in the atmosphere of the optical companion star of the LMXB GRO J1655-40 (Nova Sco 1994) observed by Israelian et al. (1998) strongly support the scenario where the BH formation was preceded by some SN-like mass ejection.

In our PNS simulations, when a BH forms with a preceding SN, the kick attributed to the proto-NS is weighted with the amount of fall back material which is equal to the difference in the final BH mass and the mass of the pre-SN iron core. We assume that all stars with an initial mass larger than 40 $M_\odot$ collapse into a BH without mass ejection. Notice however that this assumption has little effect on the results of the present paper.

**2.3. The RLOF of the primary in massive binaries: conservative or not?**

In the present section we consider the RLOF of the primary in massive binaries, i.e. the first RLOF when both components are still normal stars.

First, notice that in our PNS model we treat case A binaries in a similar way as case $B_r$ binaries. For the scope of the present paper this is more than sufficient.

It is likely that case B/C binaries with initial primary mass ≥ 40 $M_\odot$ avoid RLOF due to an LBV-type stellar wind mass-loss phase that precedes the RLOF. To calculate the evolution of these binaries we use the 'LBV scenario' as it was introduced in Vanbeveren (1991) (see also Vanbeveren et al., 1998a).

The RLOF in a case Bc/C binary with a primary mass < 40 $M_\odot$, leads to the formation of a common envelope (CE) during which it is expected that no matter is accreted by the companion star (i.e. β=0). To compute the period evolution during this phase we use the formalism of Webbink (1984) which has been adapted by de Kool (1990). In this formalism the orbital shrinkage is measured with the parameter α which is defined as the efficiency of the conversion of orbital energy into potential energy.

After more than 3 decades of extensive binary evolutionary calculations by different research teams, the following overall β-formalism for case A and case $B_r$ binaries with primary mass < 40 $M_\odot$ emerges:

- Binaries with mass ratio $q < 0.2$: spiral-in and $\beta = 0$
- Binaries with mass ratio $q > 0.4$: RLOF and $0 \leq \beta = \text{constant} = \beta_{max} \leq 1$
- Binaries with mass ratio $0.2 \leq q \leq 0.4$: we adopt a simple linear relation between 0 and $\beta_{max}$; for the scope of the present paper this is sufficient.

To calculate $\beta$ and $\beta_{max}$, one has to solve the magneto-hydrodynamic equations that describe the mass transfer, and we need a model that takes the mass accretion process into account in a realistic way. This problem is very complex and approximations are needed. An accretion model was proposed by Neo et al. (1977) but an alternative suggestion was published by Vanbeveren & De Loore (1994). In most cases the conclusions related to $\beta$ depend in a critical way on the adopted accretion model and therefore, uncertainties in the latter imply uncertainties on $\beta$. Since it can be expected that $\beta = 0$ for case $B_c$ and case C binaries, one may speculate that $\beta_{max}$ is a decreasing function of the orbital period.

We present our results for different values of $\beta_{max}$. When matter leaves the binary system we have to account for the loss of orbital angular momentum. In all our PNS results since 1997-1998, we used a formalism described by Soberman et al. (1997). Matter leaves the binary through the second Lagrangian point $L_2$ and settles in a circumbinary ring with radius $\eta A$ (with A the orbital separation). A "bare-minimum" for the circumbinary radius is found for $\eta$=1.3, which corresponds to the distance between $L_2$ and the centre of mass of the binary. However as argued by Soberman et al. this ring is unstable and is likely to fragment and to fall back on the binary components. The first stable ring corresponds to $\eta \approx 2.25$. In the present paper we calculate the variation of the binary period adopting the latter value (see also Vanbeveren et al. 1998b).

It is easy to show that significant mass loss from the binary ($\beta \leq 0.5$) is always accompanied by a large reduction of the orbital period which leads sometimes to the merging of the two components. Therefore, if PNS is computed assuming that the evolution of binaries with primary mass < 40 $M_\odot$ is highly non-conservative, we need to consider in detail the evolution of mergers as well.

## 2.4. The formation and evolution of mergers

In our PNS code we check for the merging of both components in a binary during evolution. We distinct the following cases.

### *2.4.1. Both components are normal stars*

Binaries with an initial mass ratio $q \leq 0.2$ experience a spiral-in phase during which the low mass component is dragged into the atmosphere of the most massive star and both stars merge. The final product will be a single star with a mass equal to the sum of the masses of both components, but this single merger may have a peculiar chemical composition.

Systems with a mass ratio $q > 0.2$ may merge as well due to orbital angular momentum loss during non-conservative RLOF ($\beta < 1$ and/or common envelope evolution).

The way we treat these mergers in our PNS code has been outlined in detail in Van Bever and Vanbeveren (2003).

### *2.4.2. One component is a normal star and the other is a compact star*

Obviously, the evolution of OB + NS/BH binaries is essential in order to estimate the NSM rate by PNS. It is governed by the spiral-in process during which the compact star spirals-in into the OB companion. Whether the outcome is a merged binary or not is checked by comparing the radius of the remnant helium star with its Roche radius. Our treatment in the PNS code has been outlined in Vanbeveren et al. (1998 a, b, c). It is similar to the formalism used in most of the PNS codes of other research teams (for a review, see Kalogera and Belczynski, 2001).

### *2.4.3. Both components are compact stars*

When the OB + NS/BH binary survives the spiral-in phase(s) and the SN explosion of the secondary star, a double compact star binary is formed. The further orbital evolution of the system is driven by orbital energy and angular momentum loss via gravitational wave radiation. The binary period shrinks (i.e. both components spiral-in) and the system finally merges. In our PNS simulations, the merging time $\tau$ (= the time of complete spiral-in) is calculated with the formalism of Peters (1964). Notice that $\tau$ depends on the orbital period, total mass and eccentricity of the double compact star binary at the moment of formation. The implementation in our PNS code is outlined in subsection 2.7.

### 2.5. Case BB evolution

The RLOF in a binary stops when helium starts burning in the core of the mass loser and when most of the hydrogen rich layers have been removed: the mass loser has become a hydrogen deficient CHeB star. When the post-RLOF mass is smaller than ~6 $M_{\odot}$ (corresponding with an initial mass on the ZAMS smaller than ~17 $M_{\odot}$) the further evolution deserves some attention.

Habets (1986a, 1986b) computed the evolution of helium stars with $2 \leq M/M_{\odot} \leq 4$ up to neon ignition and concluded that those with $2 \leq M/M_{\odot} \leq 2.9$ develop deep convective envelopes during the He shell burning phase, after CHeB, and expand significantly. When such a helium star is a binary member, depending on the binary period, they may fill their Roche lobe again and perform case BB RLOF. During this phase of mass transfer the star loses its remaining hydrogen layers and most of its helium layers on top of the He burning shell. The mass loss rates of the Roche lobe filling component during case BB RLOF are considerably smaller than during a case B RLOF (Dewi, 2002). This means that it is reasonable to assume that when case BB happens in a binary with a normal secondary component, the mass transfer is conservative. This is what we adopt in our PNS code.

Case BB evolution of binaries consisting of a helium star with a mass between 2 $M_{\odot}$ and 6 $M_{\odot}$ and a NS star companion has been studied in detail by Dewi et al. (2002, 2003) and by Ivanova et al. (2003). Similarly as in Habets (1986a, b) the authors illustrate the importance of the convective envelope of the donor on the mass transfer process. It is expected that in many cases a CE will be formed and the further evolution will be governed by the spiral-in process.

However, the papers listed above agree upon the fact that when certain conditions are fulfilled, a CE may be avoided during the case BB RLOF. In this latter case, when the mass transfer rate becomes larger than the critical Eddington accretion rate, the excess mass leaves the binary as a NS stellar wind with the specific orbital angular momentum of the NS. Notice however that the conditions mentioned above rely on post-CHeB stellar evolutionary calculations and uncertainties in the latter imply uncertainties in the former. The differences in the papers of Dewi et al. (2002, 2003) and of Ivanova et al. (2003) illustrate possible consequences of these uncertainties.

Dewi et al. (2002, 2003) account for stellar wind mass loss during CHeB but they use a formalism which implies that SW mass loss during CHeB is very small for a post-RLOF He star with mass ≤ 6 $M_{\odot}$ and hardly affects its CHeB and post CHeB evolution. However, this formalism is very uncertain and, to illustrate, when our equation 2 is extrapolated downwards, it cannot be excluded that at Z = 0.02 this mass loss is sufficiently large in order to suppress case BB RLOF in massive binaries, i.e. the mass that would leave the star due to case BB RLOF is lost by SW during CHeB prior to the onset of case BB. As has been outlined in Vanbeveren et al. (1998a), one of the most important differences between case BB RLOF (and the applied physics of mass transfer/mass loss from the system) versus stellar wind mass loss in relation to population synthesis is the orbital period evolution of the binary. To illustrate, when the companion is a compact star, case BB RLOF may be governed by the spiral-in process which may result in a significant hardening of the binary. The latter has a significant effect on the probability for the binary to remain bound after the second SN explosion, and thus on the birth rate of double compact star systems. Notice that when the WR mass loss rate scales with the iron abundance according to equation (2), the SW at Z = 0.002 is too small to suppress case BB RLOF.

Accounting for the discussion above, to demonstrate the importance of the assumption of case BB or stellar wind mass loss and to illustrate the effects of the physics used to describe the mass transfer during case BB, we present our simulations for the following five scenario's:

*Scenario 1*: for Z = 0.02 case BB is suppressed by stellar wind mass loss of the helium star; this stellar wind mass loss does not depend on the metallicity, which means that case BB is suppressed at Z=0.002 as well.

*Scenario 2:* for Z = 0.02 case BB is suppressed by stellar wind mass loss; this stellar wind mass loss depends on the metallicity and satisfies equation (2) which implies that case BB is not suppressed at Z=0.002. Detailed evolutionary calculations of case BB RLOF for Z=0.002 do not exist. In our PNS code we use the Z=0.02 evolutionary results of Dewi et al (2002) for Z=0.002 as well. When the companion star is a NS, we always calculate the binary period evolution by assuming that matter leaves the binary in the form of a NS stellar wind with the specific orbital angular momentum of the NS star (the resulting period evolution was discussed in Dewi et al., 2002).

*Scenario 3:* similar as scenario 2 but when the companion star is a NS, we assume that the further evolution is governed by the spiral-in process during which no significant accretion takes place on the NS (we take $\beta$=0). To compute the evolution of the binary period, we use the formalism of Webbink (1984) adapted by de Kool (1990).

*Scenario 4:* the stellar wind mass loss of hydrogen deficient CHeB stars with mass $< 6\ M_\odot$ can be neglected and thus, case BB happens independent from Z. The evolution during case BB when the companion is a NS is the same as in scenario 2.

*Scenario 5:* Similar as scenario 4 but during case BB when the companion is a NS we use the same prescription as in scenario 3.

**2.6. The binary formation rate $f_b$**

We define the parameter $f_b$ as the formation rate of binaries with the properties given at the beginning of section 2 (which corresponds in a star formation model to the fraction of binaries on the ZAMS). Remark that most of these binaries will interact, i.e. the primaries in most of these binaries will fill their Roche volume at a certain moment during their evolution.

From observational studies on spectroscopic binaries in the solar neighbourhood we know that about 33% (±13%) of the O-type stars are the primary of a massive close binary with a mass ratio q > 0.2 and a period P ≤ 100 days (Garmany et al., 1980). A similar conclusion holds for the intermediate mass B-type stars (Vanbeveren et al., 1998). Accounting for observational selection, it can be shown by binary population synthesis studies that to meet the above observations, an initial OB-type binary fraction $f_b$ larger than (50-70)% is required (Vanbeveren et al., 1997; Mason et al., 2001; van Rensbergen, 2001, and references therein).

We like to remind that in general, the binary formation rate differs from the observed overall binary fraction in a stellar population. A stellar population consists of evolved and non-evolved stars. An evolved star which is observed as a single star, can be a merged binary or it could have been a secondary of an interacting binary which was disrupted due to the SN explosion of the primary. This means that the (observed) binary fraction in a stellar population is always smaller than the real (past) binary formation rate.

### 2.7. The population of double compact star binaries predicted by PNS

Our PNS code calculates the population of NS+NS and BH+NS binaries and we investigated the effects of the different evolutionary parameters and the different input parameters in the PNS code.

In order to determine the properties of a binary after the SN explosion of one of its components, we assume that prior to the SN explosion the system was circularised. Since we treat the effects of the SN explosion on the binary parameters in full 3-D (see also Vanbeveren et al., 1998a), our PNS code is able to compute the post-SN period, space-velocity and eccentricity of the binary. The knowledge of the post-SN eccentricity is essential for computing the merging timescale (subsection 2.4.3) of the double compact star binaries.

It is not the scope of this paper to present a detailed description of the PNS predictions of the double compact star binaries. Such a discussion for the solar neighbourhood has been published by Belczynski et al. (2002) (see also Ivanova et al., 2003). We used the results in the latter papers in order to check the reliability of our PNS code (or the one of the other authors). It is worth mentioning that for the same input parameters we recover their results. We like to recall that the binary evolutionary parameters which affect most the predicted properties of the population of double compact binaries, are the average kick velocity which describes the asymmetry of the SN explosion, the energy efficiency parameter during the common envelope/spiral-in process of the OB + NS/BH binaries, the stellar wind mass loss during CHeB which decides upon the occurrence or not of case BB RLOF, and, if case BB happens, the physics of the mass transfer/mass loss from the system when the companion is a NS.

To investigate the influence of the PNS parameters on our results we will consider the following PNS models:

| Model | $\phi(q)$ | $\beta_{max}$ | $\alpha$ | $<v_{kick}>$ |
|---|---|---|---|---|
| 1 | flat | 1 | 1 | 450 |
| 2 | Hogeveen | 1 | 1 | 450 |
| 3 | Garmany | 1 | 1 | 450 |
| 4 | flat | 0.5 | 1 | 450 |
| 5 | flat | 1 | 0.5 | 450 |
| 6 | flat | 1 | 1 | 150 |

**Table 1**. The different PNS models for which are simulations are made.

## 3. The chemical evolutionary model including binaries and neutron star mergers

The NSM rate in galaxies depends on the physics of Galaxy formation, on the overall star formation rate and on stellar evolution. At least the latter depends on the metallicity. Therefore, to calculate the temporal evolution of the NMS rate of galaxies, it is essential to combine a star formation model (SFM), a galactic chemical evolutionary model (CEM) and a PNS model (notice that in general, a CEM includes a SFM but since the NSM rate depends critically on the SFM we will consider it separately). The NSM rate depends on the binary population and therefore, to be consistent, also the CEM has to account for the evolution of binaries and their chemical yields. The Brussels CEM that accounts for the evolution of binaries has been described in De Donder and Vanbeveren (2002). Our CEM uses the galaxy and star formation model of Chiappini et al. (1997) [see also Talbot and Arnett (1975) and Chiosi (1980)].

We explored the effects of binaries on the overall SFM and concluded that, although interacting binaries return less matter to the interstellar medium (due to a higher formation rate of NSs and BHs), the effects of binaries is very small even for a constant binary frequency of 70%. The theoretical predicted total Galactic SFR (assuming a Galactic radius of 18 kpc) is given in figure 3 and should be typical for all spiral galaxies which form in two phases of major infall discussed in the papers cited above. The corresponding present massive star formation rate is ~$2{\cdot}10^{-2}$ $yr^{-1}$.

For the iron SN yields we use the values of WW95 reduced with a factor of two. It was shown by Timmes et al. (1995) that the reduced iron yields give a better fit with the observed abundance evolution and better correspond with the observed iron abundance in SN1987A and SN1993J (Thomas et al. 1998). The predicted time evolution of the iron content (relative to hydrogen or [Fe/H]) is given in figure 4.

For the r-process yields we account only for the contribution from NSMs. We assume for simplicity that merging NS+NS and BH+NS systems eject the same amount of matter independent from the total mass of the system and that all of the ejecta is r-process material. Numerical computations of an NSM event by Rosswog et al. (1999) show that a total mass $M_{ej}$ from $4.10^{-3}$ to $4.10^{-2}$ $M_{\odot}$ can be ejected. In our CEM simulations we use both limiting values. Notice however that Rosswog et al. apply Newtonian physics and that the inclusion of general relativity in the theory results into much smaller ejecta (Oechslin and Thielemann, 2001). For the solar r-abundance we use $X_{r,\odot} \approx 10^{-7}$ (Käppeler et al., 1989).

Remind that in order to calculate the temporal evolution of the NSM rate and of the corresponding r-process yields, in the CEM one has to account for the total lifetime of the binary system which is given by the formation time scale of the double compact star binary plus its merging timescale.

**4. Results.**

Figures 5 and 6 illustrate the time evolution of the merger rate of NS+NS and BH+NS binaries computed for the different case BB scenarios (discussed in section 2.5). All the computations are made with a constant binary frequency of 70% (on the ZAMS) during the whole Galactic evolution and with PNS model 1. Since we follow in detail the chemical evolution of the galaxy, the time evolution of the merger rate is given as a function of the metallicity. We conclude that,

- The moment during Galactic evolution at which the first merging NS+NS and BH+NS pairs appear, depends primarily on whether or not case BB RLOF occurs in massive binaries and on the physics used to describe the latter, i.e. CE evolution or isotropic mass loss via a SW from the NS.
- Pairs of merging BH+NS form earlier than NS+NS pairs. The BHs in the former systems received a smaller kick velocity during the SN explosion and have on average smaller post-SN orbital periods than NS+NS pairs.
- The overall temporal behaviour of the merger rate is typical for the adopted SFR i.e. a rapid increase, followed by a plateau with a decline towards [Fe/H]=0. The knick at [Fe/H] ~ -0.5 is caused by the sudden decrease in the SFR at t=2 Gyr that roughly corresponds to the end of the thick disk phase.
- The predicted present NSM rate is between ~$10^{-6}$ and $10^{-4}$/yr and in agreement with the observational estimated rate (e.g. De Donder and Vanbeveren, 1998; Kalogera and Belczynski, 2001).

To illustrate the influence of the PNS model parameters, figures 7 and 8 show the results for the different PNS models (table 1). They are computed with scenario 5 for case BB RLOF. As expected, the average kick magnitude and the spiral in efficiency parameter of OB+NS/BH binaries are critical for the predicted rates (in particular for the NS+NS rates).

Figures 9 and 10 show the time evolution of the abundance ratio [Eu/Fe] versus [Fe/H] (predicted for the solar neighbourhood) which is representative for the evolutionary behaviour of the r-process elements produced by merging NS+NS and BH+NS binaries. The results for the other PNS models in combination with the case BB scenario 5, are given in figures 11 and 12.

These results illustrate the following important overall conclusion.

***Overall conclusion***

The binary neutron star merger model can explain the r-process enrichment of the Galaxy. However, this conclusion depends critically on

- the effects on the orbital binary parameters of an asymmetric SN explosion
- the physics of CE evolution
- the adopted scenario and physics of case BB evolution
- the amount of ejected r-process material during the merger event

**References**

Andreani, P., Vangioni-Flam, E., Audouze, J., 1988, ApJ., 334, 698-706.

Baumgarte, T., Cook, G., Scheel, M., Shapiro, S., Teukolsky, S., 1997, Phys. Rev. Lett. 79, 1182.

Belczynski, K., Kalogera, V., Bulik, T., 2002, ApJ, 572, 407.

Burbridge, E.M., Burbridge, G.R., Fowler, W.A., Hoyle, F., 1957, Rev. Mod. Phys., 29, 547.

Burris, D.L., Pilachoski, C.A., Armandroff, T.E., Sneden, C., Cowan, J., Roe, H., 2000, ApJ., 544, 302-319.

Cameron, A.G.W., 1957, Atomic Energy of Canada, Ltd., CRL-41.

Cayrel, R., Hill, V., Beers, T.C., et al., 2001, Nature, 409, 691.

Chiappini, C., Matteucci, F., Gratton, R., 1997, ApJ., 477, 765.

Chiosi, C., A&A., 83, 206-216.

Cowan, J.J., Thielemann, F.-K., Truran, J.W., 1991, Phys. Rep., 208, 267.

Davies, M., Benz, W., Piran, T., Thielemann, F.-K., 1994, ApJ., 431, 742.

De Donder, E., Vanbeveren, D., 1998, A&A, 333, 557.

De Donder, E., Vanbeveren, D., 2002, NewA, 7, 55.

De Donder, E., Vanbeveren, D., 2003, A&A., submitted.

de Kool, M.: 1990, ApJ 358, 189.

Dewi, J. D. M., Pols, O. R., Savonije, G. J., van den Heuvel, E. P. J., 2002, MNRAS, 331, 1027-1040.

Dewi, J. D. M., Pols, O. R., 2003, MNRAS, submitted.

Edvardsson, B., Andersen, J., Gustafsson, B., Lambert, D. L., Nissen, P. E., Tomkin, J., 1993, A&A, 275, 101.

Freiburghaus, C.,Rembges, J., Rauscher, T., et al., 1999a, ApJ., 516, 381.

Freiburghaus, C., Rosswog, S., Thielemann, F.-K., 1999b, ApJ., 525, L121.

Garmany, C.D., Conti, P.S., Massey, P., 1980, ApJ., 242, 1063.

Habets, G.M.H.J., 1986a, A&A., 165, 95.

Habets, G.M.H.J., 1986b, A&A., 167, 61.

Hamann, W.-R., Koesterke, L., Wessolowski, U., 1995, A&A., 299, 151.

Hill, V., Plez, B., et al., 2002, A&A., 387, 560-579.

Hoffman, R.D., Woosley, S.E., Qian, Y.-Z., 1997, ApJ., 482, 951.

Hogeveen, S. J., 1991, "The Mass Ratio Distribution of Binary Stars", PhD. Thesis.

Hogeveen, S.J., 1992, Ap&SS., 196, 299.

Humphreys, R.M., McElroy, D.B., 1984, ApJ., 284, 565-577.

Humphreys, R.M., Davidson, K., 1994, PASP, 106, 1025.

Ishimaru, Y., Wanajo, S., 1999, ApJ., 511, L33-L36.

Israelian, G., Garcia, R.J., Rebolo, R., 1998, ApJ, 507, 805.

Ivanova, N., Belczynski, K., Kalogera, V., Rasio, F.A., Taam, R.E., 2002, astro-ph/0210267.

Janka, H.T., Ruffert, M., 1996, A&A., 307, L33.

Jura, M., 1987, ApJ., 313, 743.

Kalogera, V., Belczinsky, K., 2001, in "The Influence of Binaries on Stellar Population Studies", 2001, p.447, ed. D. Vanbeveren, Kluwer Academic Publishers.

Käppeler, F., Beer, H., Wishhak, K., 1989, Rep. Prog. Phys., 52, 945.

Kippenhahn, R., Weigert, A., 1967, Z. Astrophys. 65, 251.

Lauterborn, D., 1970, A&A., 7, 150.

Lorimer, D.R., Bailes, M., Harrison, P.A., 1997, MNRAS, 289, 592.

Mason,B.D., Gies, D.R., Hartkopf, W.I., 2001, in "The Influence of Binaries on Stellar Population Studies", 2001, p.37, ed. D. Vanbeveren, Kluwer Academic Publishers.

Mathews, G.J., Bazan, G., Cowan, J.J., 1992, ApJ., 391, 719-735.

Meyer, B.S., McLaughlin, G.C., Fuller, G.M., 1998, Phys. Rev., C, 58, 3696.

Nelemans, G., Tauris, T.M., van den Heuvel, E.P.J., 1999, A&A., 352, L87-L90.

Neo, S., Miyaji, S., Nomoto, K., Sugimoto, D., 1977, Publ. Astron. Soc. Japan, 29, 249.

Nugis, T., Lamers, H.J.G.L.M., 2000, A&A, 360, 227.

Oechslin, R., Thielemann, F.K., 2001, in "The Influence of Binaries on Stellar Population Studies", 2001, p.37, ed. D. Vanbeveren, Kluwer Academic Publishers.

Pagel, B.E.J., Tautvasiene, G., 1997, MNRAS, 288, 108.

Peters, P.C., Phys. Rev., 136, B1224.

Popova, E.I., Tutukov, A.V., Yungelson, L.R., 1982, Astron. Space. Sci., 88, 55.

Reid, N., Tinney, C., Moud, J., 1990, ApJ, 348, 98-119.

Rosswog, S., Liebendorfer, M., Thielemann, F.-K., Davies, M., Benz, W., Piran, T., 1999, A&A., 341, 499.

Rosswog, S., Davies, M.B., Thielemann, F.-K., Piran, T., 2000, A&A., 360, 171.

Rosswog, S., Freiburghaus, C., Thielemann, F.-K., Davies, M.B., 2001, 20th Texas Symposium on relativistic astrophysics, Austin, Texas, 10-15 December 2000, Melville, NY: American Institute of Physics, 2001, xix, 938 p. AIP conference proceedings, Vol. 586, eds; J. Craig Wheeler and Hugo Martel.

Ruffert, M., Janka, H.-T., 1998, A&A., 338, 535.

Salpeter, E.E., 1955, ApJ., 121, 161.

Shetrone, M.D., 1996, AJ., 112, 1517.

Sneden, C., Cowan, J., Ivans, I.I., Fuller, G.M., Burles, S., Beers, T.C., Lawler, J.E., 2000, ApJ., 533:L139-142.

Soberman, G.E., Phinney, E.S., Van den Heuvel, E.P.J., 1997, A&A., 327, 620-635.

Suess, H.E., Urey, H.C., 1956, Rev. Mod. Phys., 28, 53.

Takahashi, K., Witti, J., Janka, H.-Th, 1994, A&A., 286, 857.

Talbot, R.J., Jr., Arnett, W.D., 1975, AJ., 197, 551-570.

Thomas, D., Greggio, L., Bender, R., 1998, MNRAS, 296, 119.

Timmes, F.X., Woosley, S.E., Weaver, T.A., 1995, ApJSS., 98, 617.

Travaglio, C., Galli, D., Gallino, R., Busso, M., Ferrini, F., Straniero, O., 1999, ApJ., 521, 691-702.

Travaglio, C., Galli, D., Burkert, A., 2001, ApJ., 547, 217-230.

Qian, Y.-Z., Woosley, S.E., 1996, ApJ., 471, 331.

Qian, Y.-Z., 2000, ApJ., 534, L67-L70.

Van Bever, J., Vanbeveren, D., 2003, A&A., in press.

Van Rensbergen, W., 2001, in "The Influence of Binaries on Stellar Population Studies", 2001, p.21, ed. D. Vanbeveren, Kluwer Academic Publishers.

Vanbeveren, D., 1991, A&A., 252, 159.

Vambeveren, D., De Loore, C., 1994, A&A., 290, 129.

Vanbeveren, D., Van Bever, J., De Donder, E., 1997, A&A., 317, 487-502.

Vanbeveren, D., Van Rensbergen, W., De Loore, C., 1998a, ARA&A 9, 63.

Vanbeveren, D., Van Rensbergen, W., De Loore, C., 1998b, monograph The Brightest Binaries, eds. Kluwer Academic Publishers: Dordrecht.

Vanbeveren, D., De Donder, E., Van Bever, J., Van Rensbergen, W., De Loore, C., 1998c, NewA 3, 443.

Vereshchagin, S., Kraicheva, Z., Popova, E., Tutukov, A.V., Yungel'son, L., 1987, Pis ma Astronomicheskii Zhurnal, 13, 63.

Vereshchagin, S., Tutukov, A.V., Yungel'son, L., Kraicheva, Z., Popova, E., 1988, Ap&SS, 144, 245.

Webbink, R.F., 1984, ApJ., 277, 355.

Woolf, V.M., Tomkin, J., Lambert, D.L., 1995, ApJ., 453, 660-672.

Woosley, S.E., Wilson, J.R., Mathews, G.J., Hoffman, R.D., Meyer, B.S., 1994, ApJ., 433, 229.

Woosley, S. E., Weaver, T. A., 1995, ApJ.SS., 101.

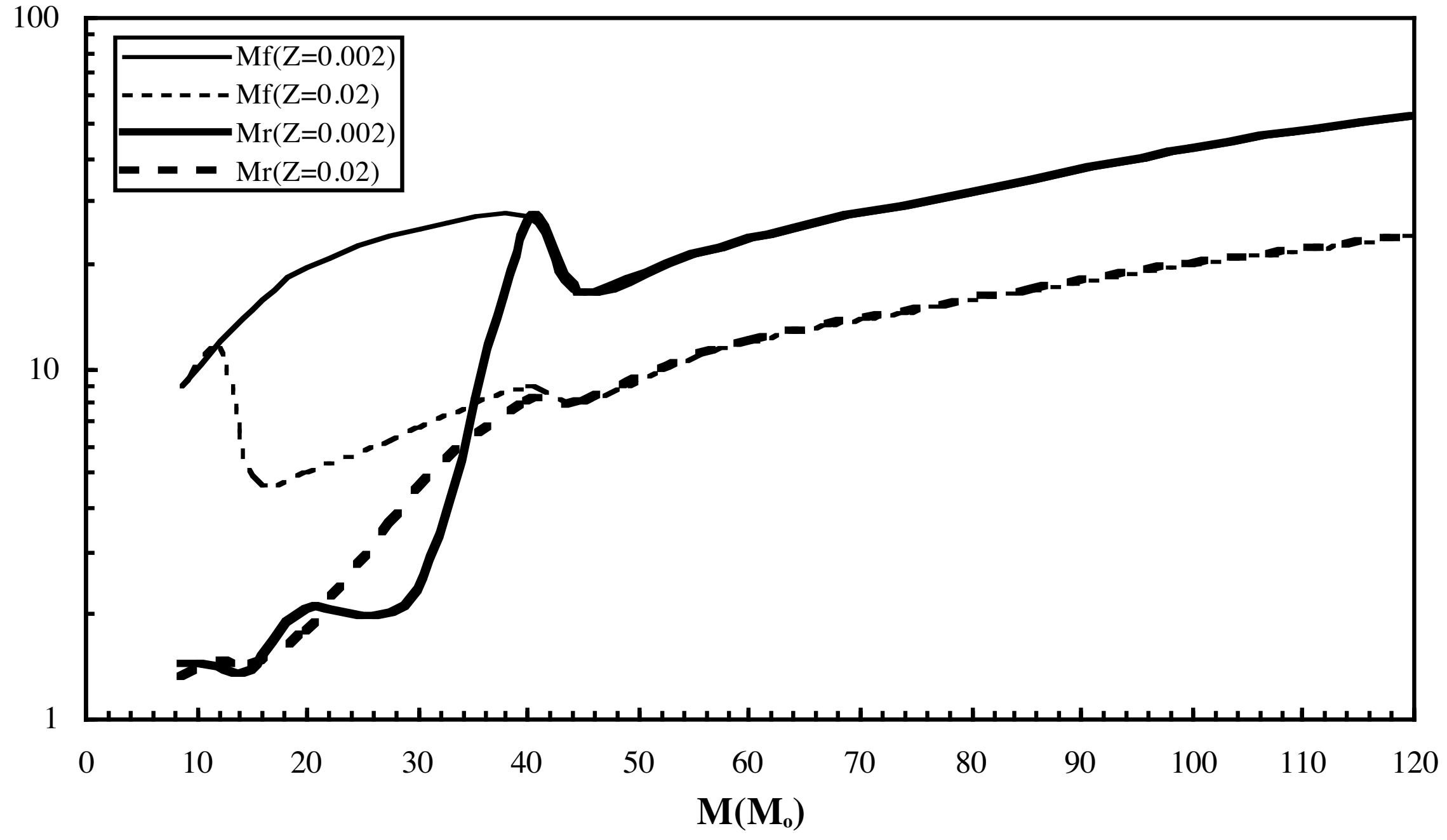

**Figure 1:** The pre core collapse ($M_f$) and post core collapse ($M_r$) masses of massive single stars for Z=0.02 and Z=0.002.

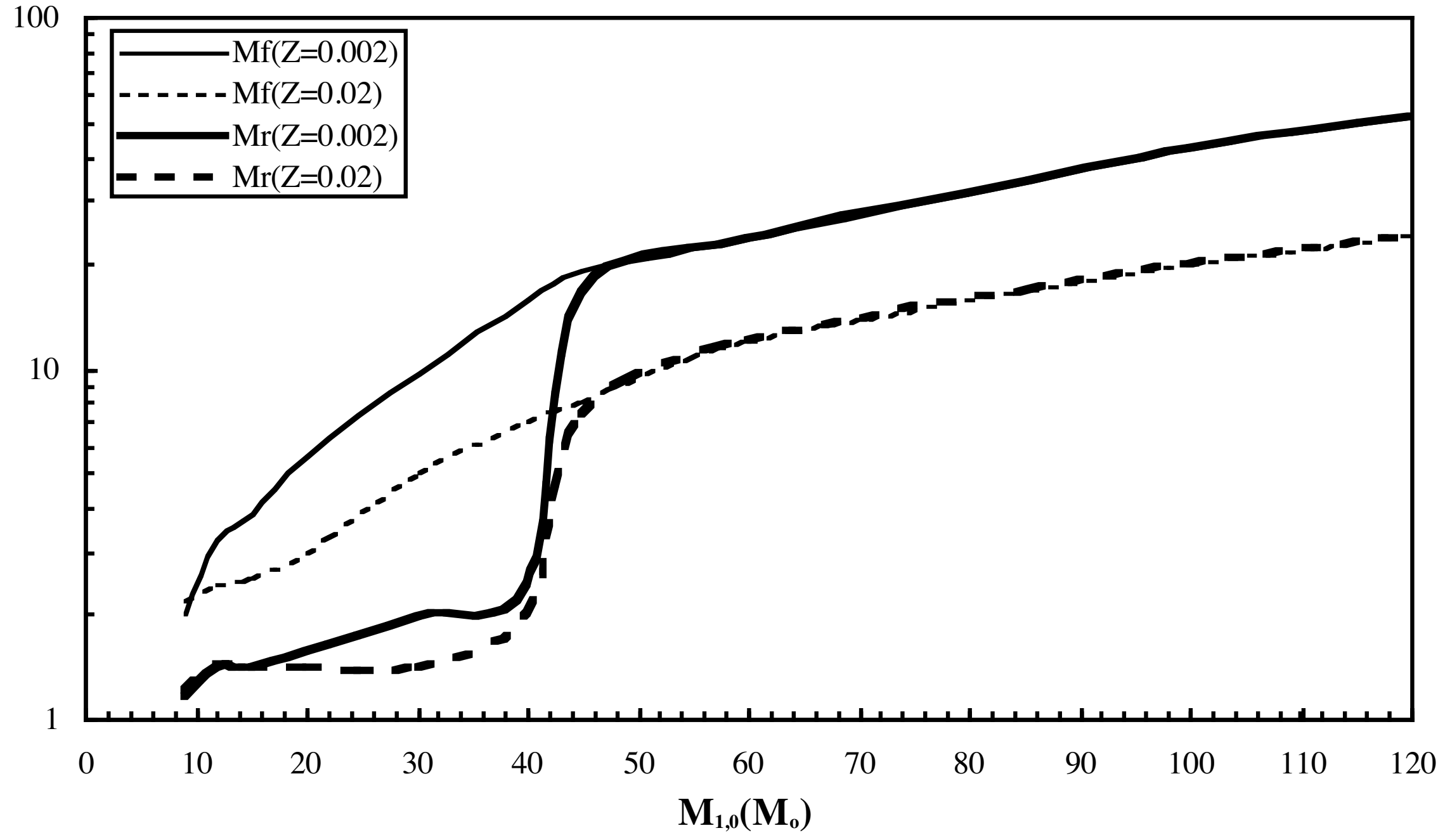

**Figure 2:** The pre core collapse ($M_f$) and post core collapse ($M_r$) masses of primary stars for Z=0.02 and Z=0.002.

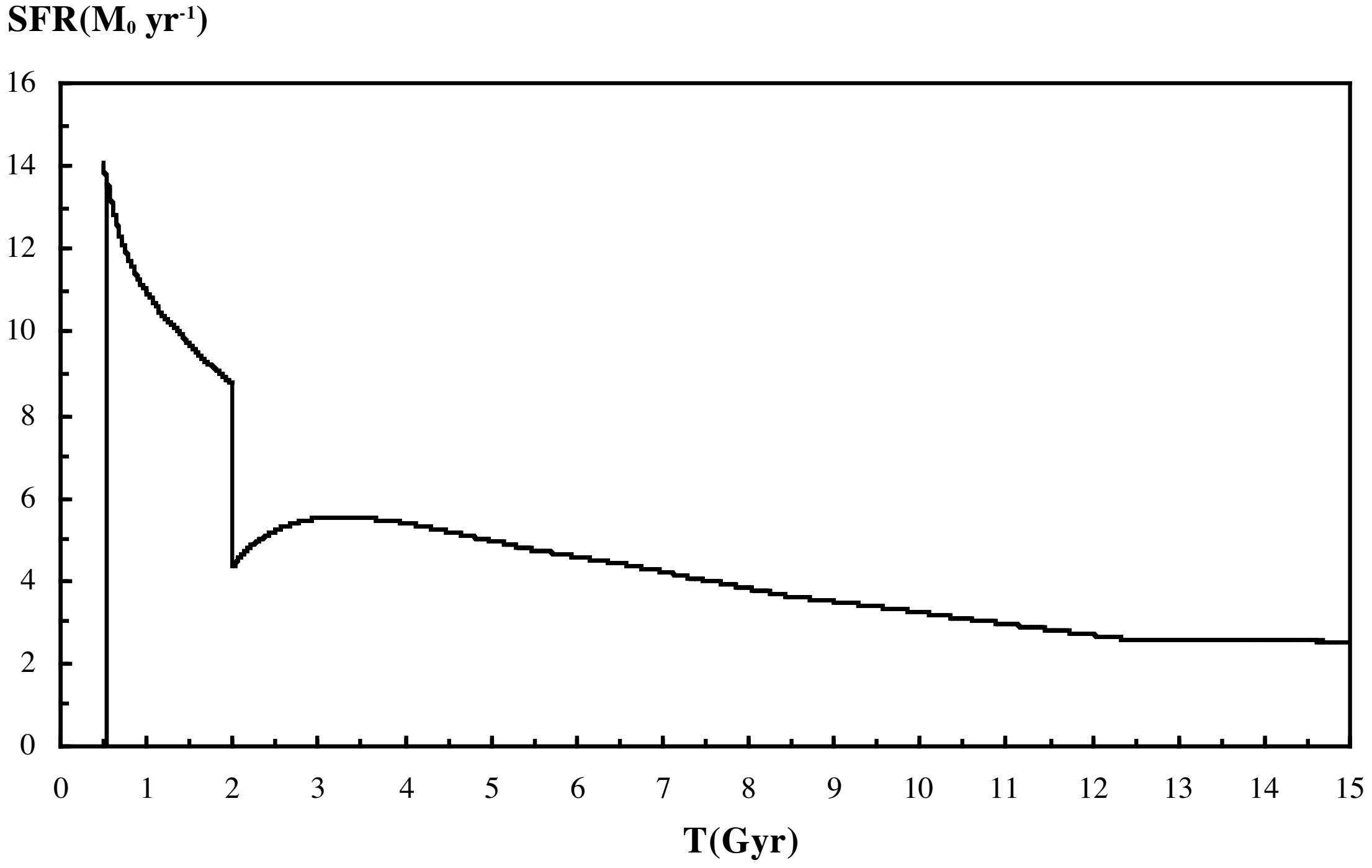

**Figure 3:** The total Galactic star formation rate as a function of time.

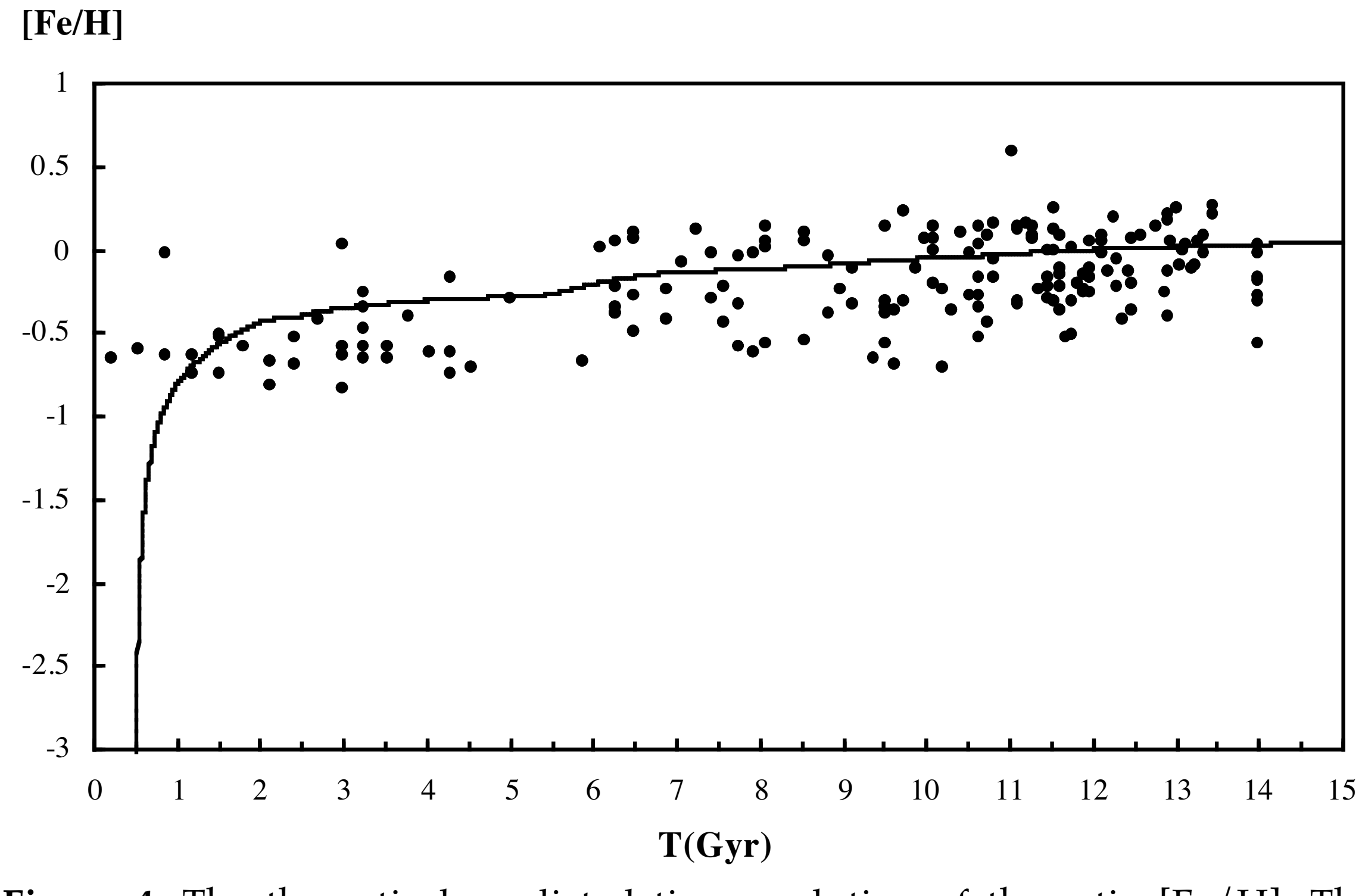

**Figure 4:** The theoretical predicted time evolution of the ratio [Fe/H]. The observational data points are from Edvardsson et al. (1993).

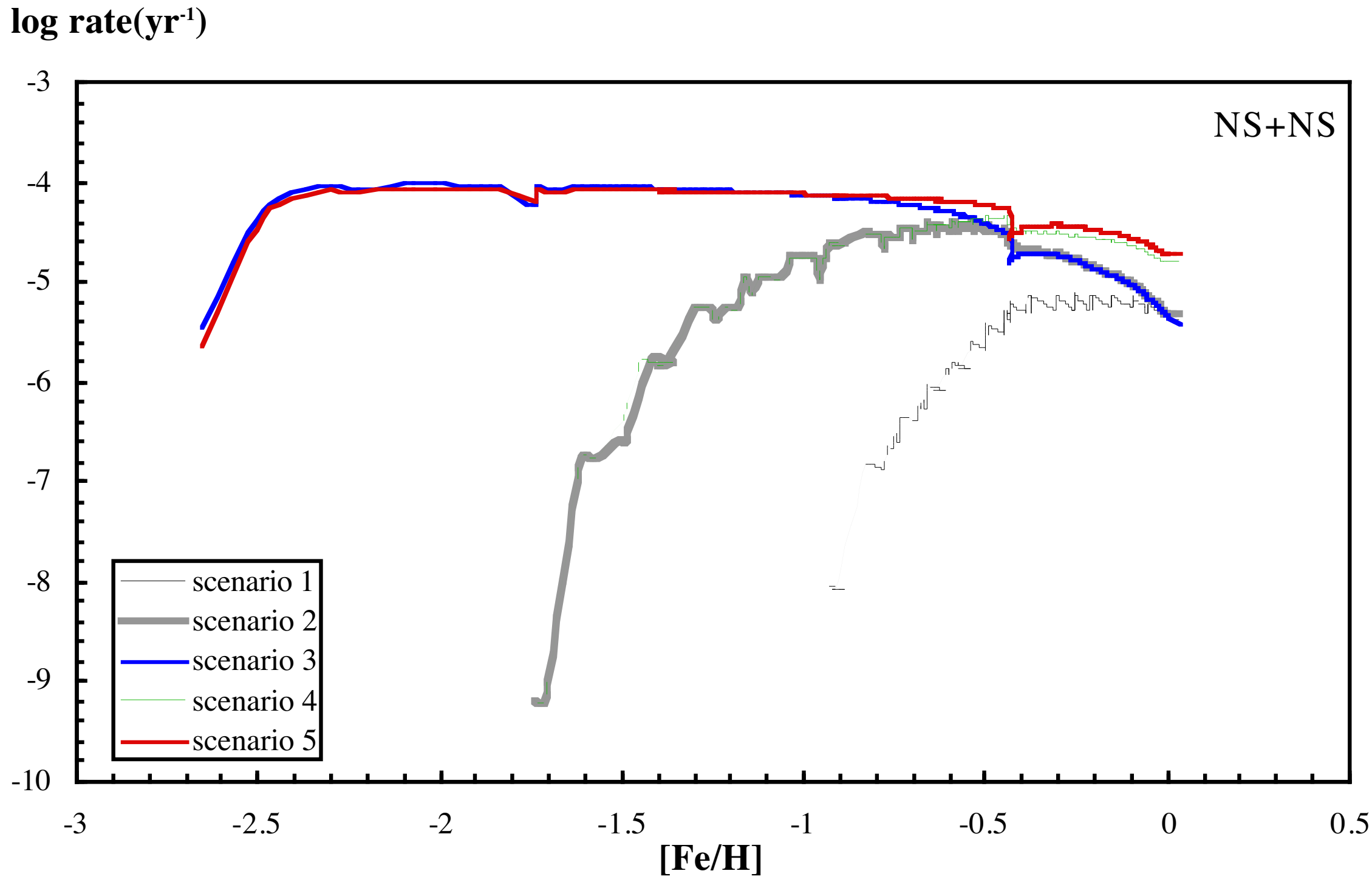

**Figure 5:** The time evolution of the Galactic merger rate of NS+NS pairs as predicted by the CEM for the different case BB scenario's and for PNS model 1.

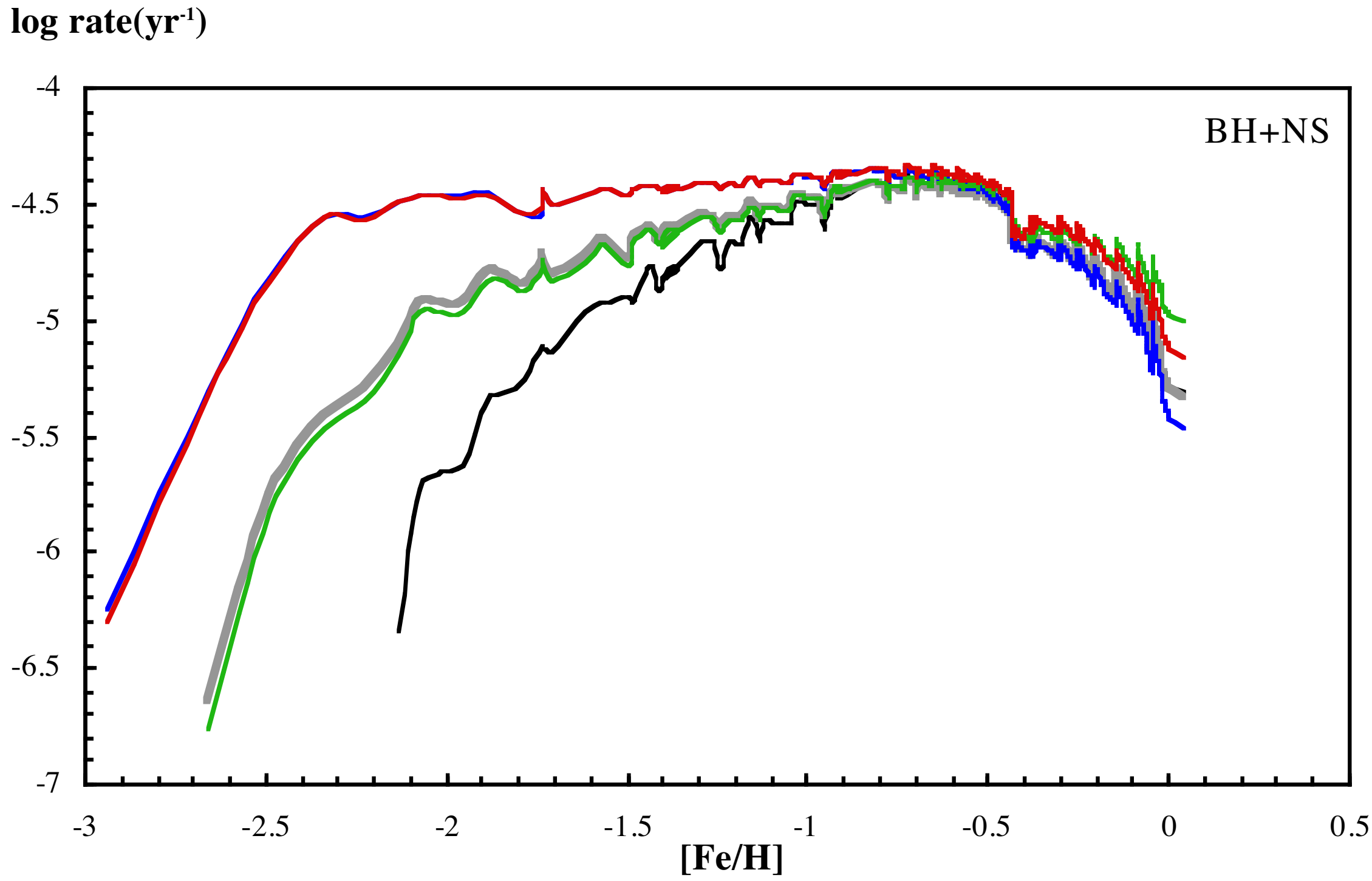

**Figure 6:** The time evolution of the Galactic merger rate of BH+NS pairs as predicted by the CEM for the different case BB scenario's and for PNS model 1.

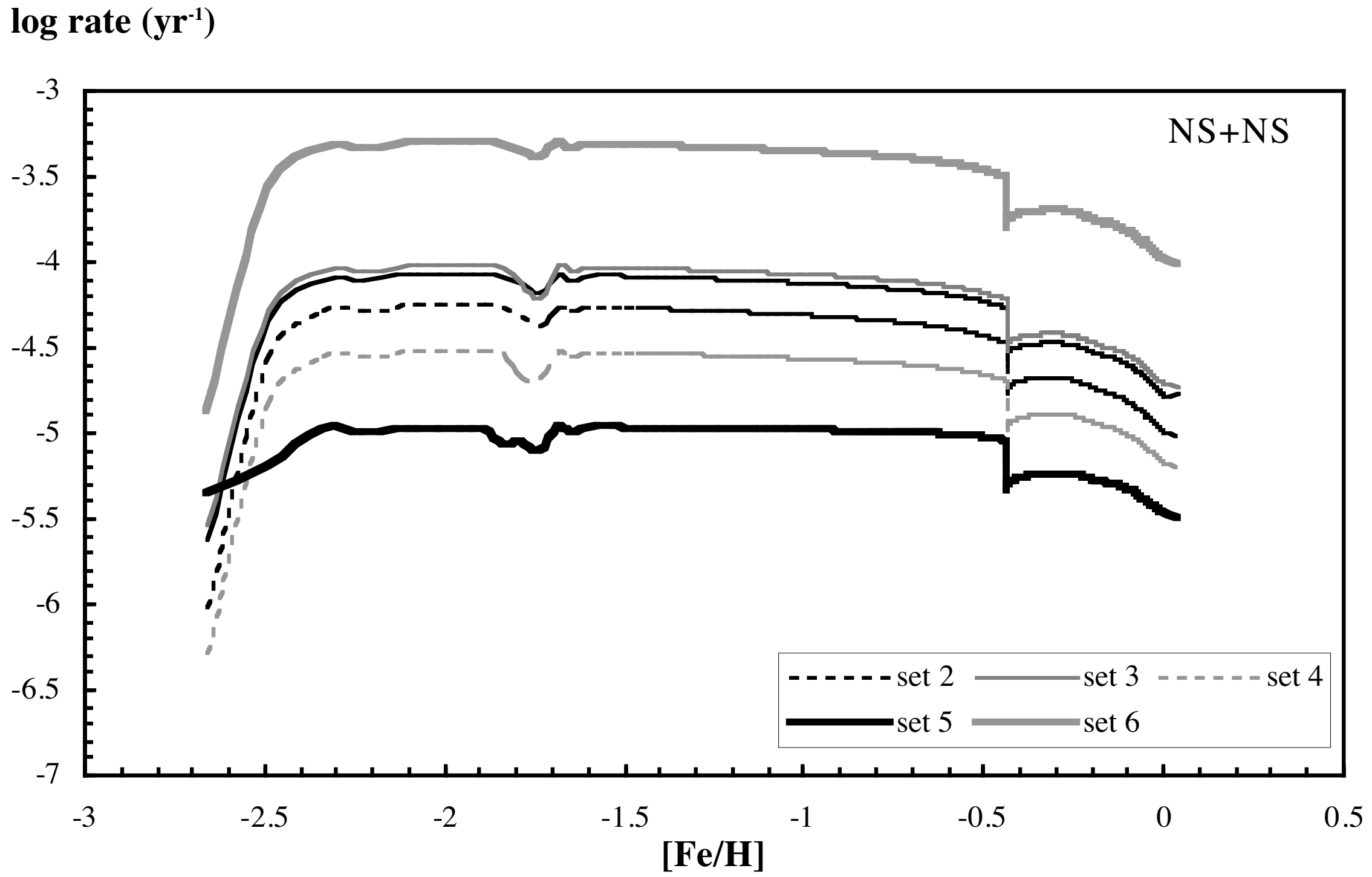

**Figure 7:** The time evolution of the Galactic merger rate of NS+NS pairs as predicted by the CEM for the different PNS models and case BB scenario 5.

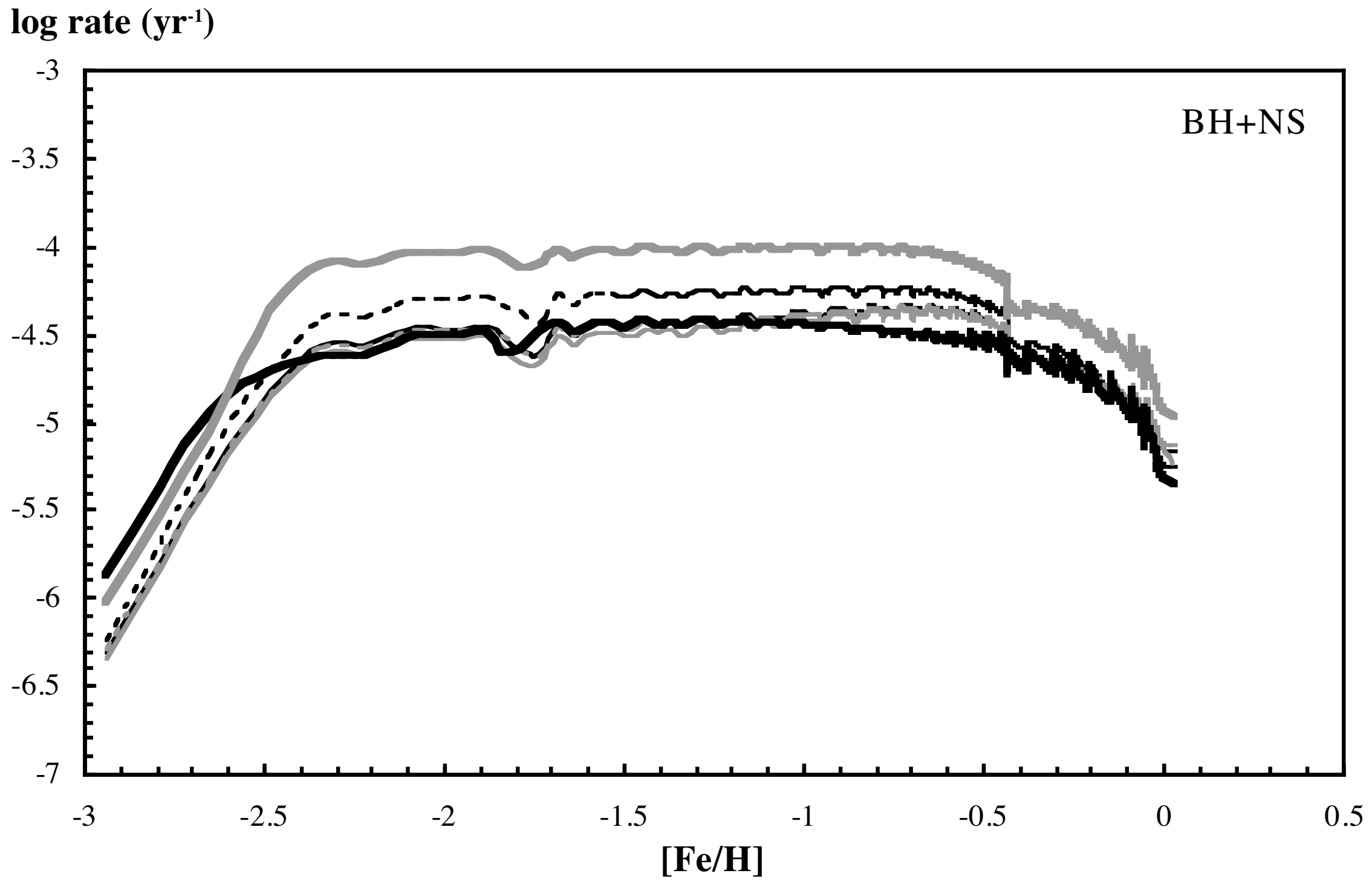

**Figure 8:** The time evolution of the Galactic merger rate of BH+NS pairs as predicted by the CEM for the different PNS models and case BB scenario 5.

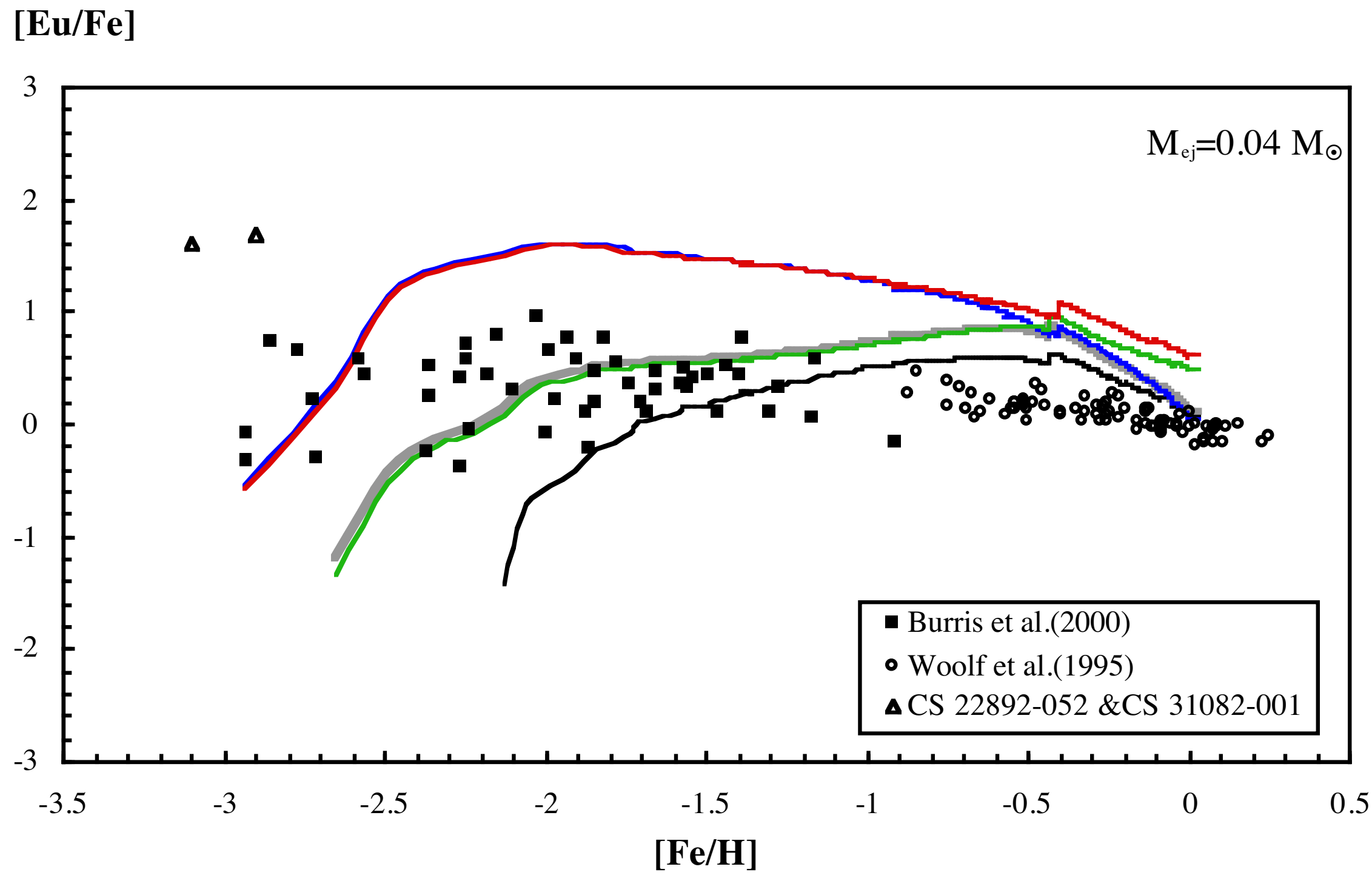

**Figure 9:** The theoretically predicted [Eu/Fe] vs. [Fe/H] relation (for the solar neighborhood) predicted for the different case BB scenarios and $M_{ej}=0.04M_{\odot}$. The observational data are from the sources given in the legend.

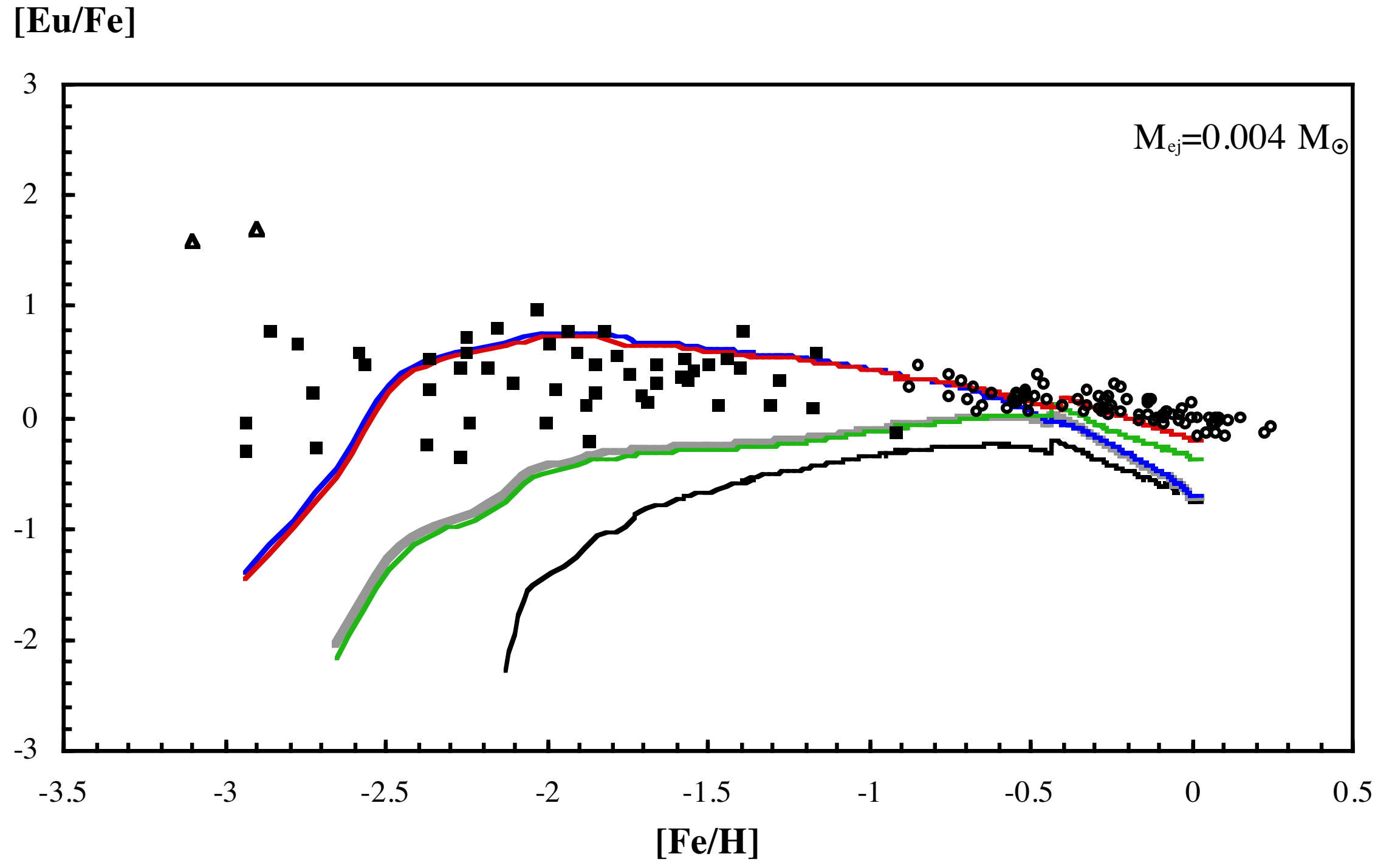

**Figure 10:** The same as figure 9 but for $M_{ej}=0.004M_{\odot}$.

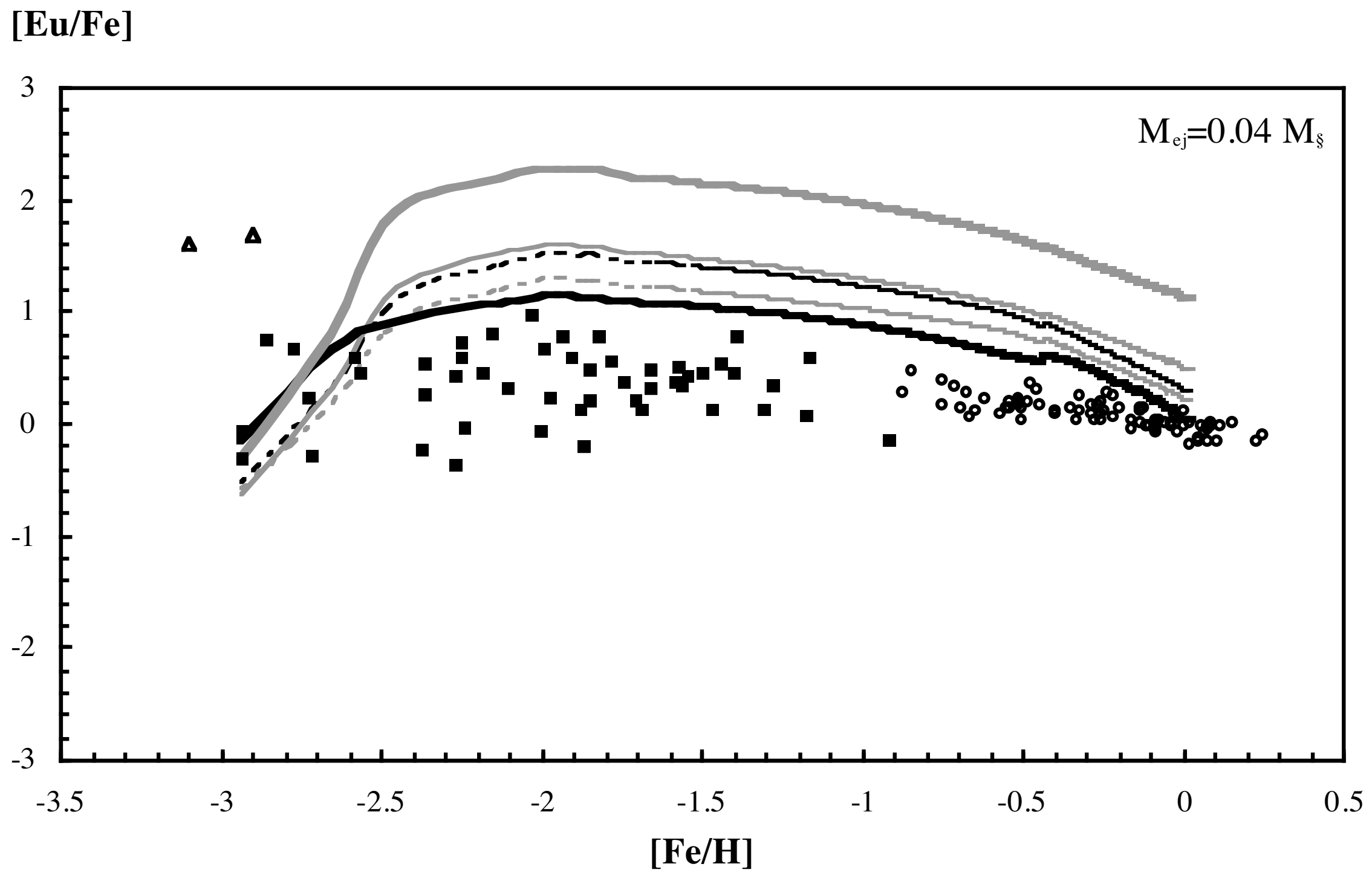

**Figure 11:** The theoretically predicted [Eu/Fe] vs. [Fe/H] relation computed with the different PNS models, case BB scenario 5 and $M_{ej}$=0.04$M_{\odot}$.

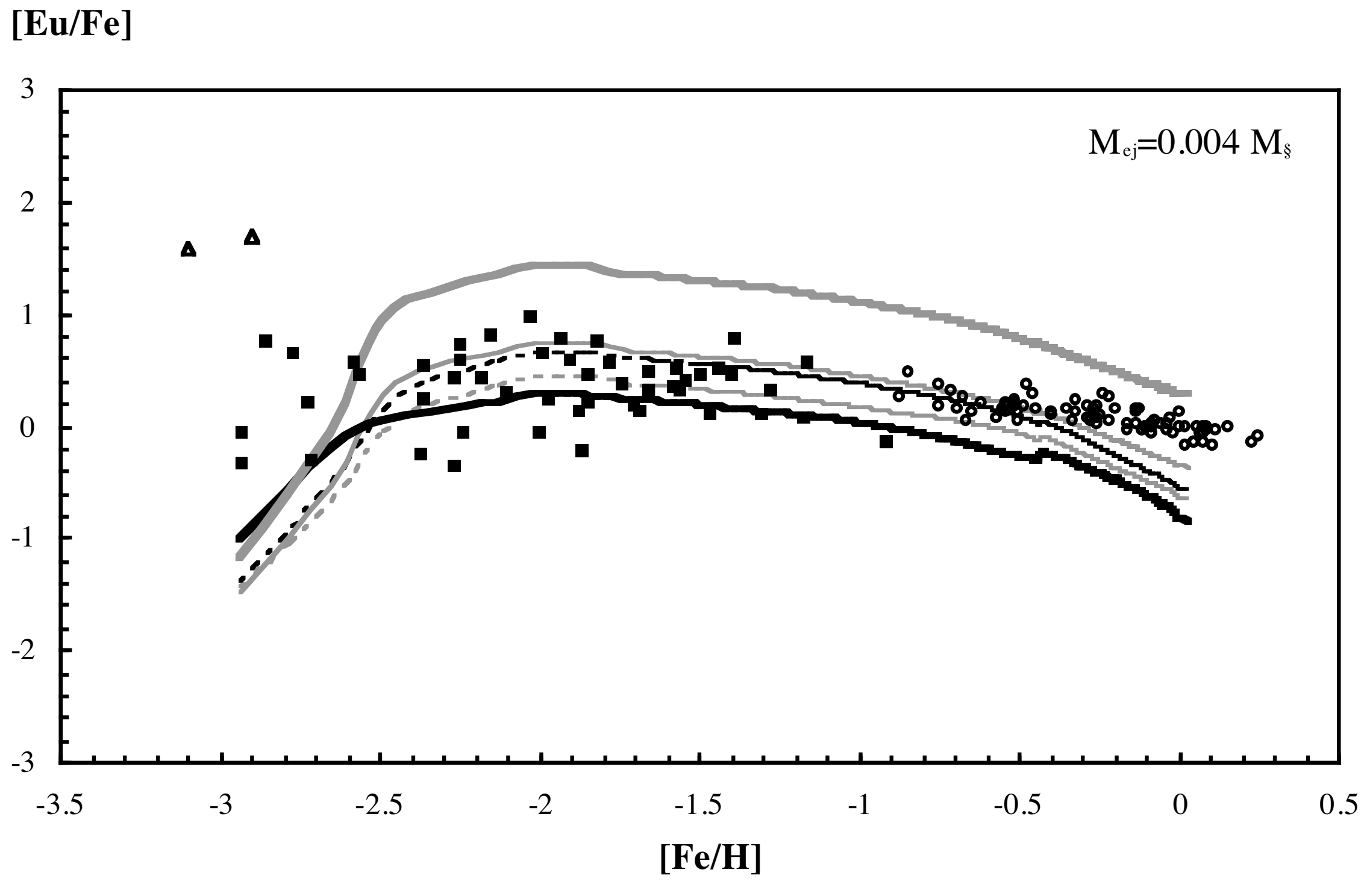

**Figure 12:** The theoretically predicted [Eu/Fe] vs. [Fe/H] relation computed with the different PNS models, case BB scenario 5 and $M_{ej}$=0.004$M_{\odot}$.